\documentclass{PoS}

\usepackage[utf8]{inputenc}
\usepackage[T1]{fontenc}
\usepackage{amsmath,amsthm, amssymb, latexsym}
\usepackage{slashed}

\usepackage{latexsym}
\usepackage{subfigure}
\usepackage{multicol}
\usepackage{multirow}
\usepackage{mathrsfs}
\usepackage{verbatim}
\usepackage{latexsym}
\usepackage{subfigure}
\usepackage{multicol}

\usepackage{amsmath}

\usepackage{algorithmic}

\usepackage{array}

\usepackage{graphicx}

\usepackage[absolute,overlay]{textpos}

\title{Implementation of the conjugate gradient algorithm
on FPGA devices}

\ShortTitle{Implementation of the conjugate gradient algorithm
on FPGA devices}

\author{\speaker{Piotr Korcyl}\\
        M. Smoluchowski Institute of Physics, Jagiellonian University, ul. \L ojasiewicza 11, 30-348 Krak\'ow, Poland\\
        Institut f\"ur Theoretische Physik, Universit\"at Regensburg, 93040 Regensburg, Germany\\
        E-mail: \email{piotr.korcyl@uj.edu.pl}}

\author{Grzegorz Korcyl \\
        Department of Information Technologies,  Faculty of Physics, Astronomy and Applied Computer Science, Jagiellonian University, ul. \L ojasiewicza 11, 30-348 Krak\'ow, Poland\\
        E-mail: \email{grzegorz.korcyl@uj.edu.pl}}

\abstract{
Results of porting parts of the Lattice Quantum Chromodynamics code to modern FPGA devices are presented. A single-node, double precision
implementation of the Conjugate Gradient algorithm is used to invert numerically the Dirac-Wilson operator on a 4-dimensional grid on a Xilinx Zynq
evaluation board. The code is divided into two software/hardware parts in such a way that the entire multiplication by the Dirac operator is
performed in programmable logic, and the rest of the algorithm runs on the ARM cores. Optimized data blocks are used to efficiently use data
movement infrastructure allowing to reach intervals of 1 clock cycle. We show that the FPGA implementation can offer a comparable performance compared
to that obtained using Intel Xeon Phi KNL.
}

\FullConference{The 36th Annual International Symposium on Lattice Field Theory - LATTICE2018\\
		22-28 July, 2018\\
		Michigan State University, East Lansing, Michigan, USA.}

\begin{document}

\section{Introduction}

Field Programmable Gate Array (FPGA) devices enjoy more and more attention from the HPC perspective in the last couple of years. Several FPGA clusters are already in operation, constructed along two different architectural philosophies: either the FPGA devices are considered as the main and only compute resource \cite{aruz} or they are used as hardware accelerators attached to some CPU \cite{Kenter_2016}. The advantages and disadvantages of both scenarios depend heavily on the memory access patterns in the given problem being solved. In either case, the tremendous increase of available of resources in a single FPGA device witnessed recently and new design methodologies yield a powerful candidate for HPC systems. Several features of FPGA are especially attractive: natural parallelism allows to allocate many instances of a given logical component with all of them running in parallel; multi-gigabit transceivers which are embedded within the FPGA package allow for a communication between several FPGA devices circumventing any additional hardware layer; no dependency nor overhead imposed by the operating system; and last but not least logic resources which can be dynamically reconfigured to accelerate another part of the algorithm. All this comes with a power consumption two orders of magnitude smaller than current HPC compute node processors.

In this contribution we benchmark the most compute-intensive kernel in Lattice QCD, namely the Dirac matrix multiplication on a single FPGA device. To this end we implement a simple Conjugate Gradient solver and use it to invert the Dirac matrix on a small lattice using a single Xilinx ZCU102 platform. The latter possess an 4-core ARM processor which we use to run the main program function and a programmable logic block where we implement the hardware-accelerated Dirac matrix multiplication. In these proceedings we briefly describe the conditions which must be fulfilled in order to fully pipeline the computations of all stencils. We also investigate the limitations on the possible lattice sizes that can be analyzed on a single FPGA.
Along the way we briefly describe modern tools which allow the user to program solely in a high-level programming language such as c++ thus considerably simplifying the effort to  program the FPGA devices.

The results discussed in this contribution were described in detail in Ref. \cite{Korcyl:2018pjc}. Previous, independent efforts by other groups going in the same direction were presented at conferences in 2005 \cite{4100953} and 2006 \cite{DBLP:conf/ipps/CallananNOSG05} by O. Callanan and recently in 2016 by T. Janson \cite{Janson2017HighlyPL}.

In the following section \ref{sec. lattice} we start by very briefly describing our benchmark problem, the Dirac operator and the Conjugate Gradient algorithm, then we discuss the FPGA technology, concentrating on their possible advantages in the context of HPC in section. The results of our benchmark tests are discuss in section \ref{sec. benchmarks}, before the conclusions.

\section{Implementation details}
\label{sec. lattice}

The Dirac operator matrix acting on a spinor vector is defined as \cite{gattringer}
\begin{multline}
D(m,n)^{AB}_{\alpha \beta} \psi_{\beta}^B(n) = (m_q + 4) \psi_{\alpha}^A(n) +\\+
 \frac{1}{2} \sum_{\mu=0}^3 \Big[ U_{\mu}^{AB}(n) P^{-\mu}_{\alpha \beta} \psi_{\beta}^B(n+\hat{\mu}) +
U^{\dagger, AB}(n-\hat{\mu}) P^{+\mu}_{\alpha \beta} \psi_{\beta}^B(n-\hat{\mu}) \Big]
\label{eq. dirac op}
\end{multline}
with the convention that two indices are always summed over, and 
$P^{\pm \mu} = 1 \pm \gamma_{\mu}$.
We exploit the fact that due to the specific structure of the $P^{\pm \mu}$ matrices, the two lower $\alpha$ components of the spinor are related to the upper two by a simple
rescaling \cite{tadonki}, hence one can halve the number of multiplications by the $U$ matrix. 

The problem we are addressing is that of finding the solution $\psi_B^{\beta}(m)$ given the rght-hand side vector $\eta_A^{\alpha}(n)$,
\begin{equation}
D^{AB}_{\alpha \beta}(n,m) \psi_B^{\beta}(m) = \eta_A^{\alpha}(n).
\label{eq. dirac eq}
\end{equation}
The Dirac operator as defined in Eq. \ref{eq. dirac op} 
satisfies the $\gamma_5$-hermiticity, $\gamma_5 D \gamma_5 = D^{\dagger}$, so in order to apply the CG algorithm one needs to solve for $D D^{\dagger}$ which is hermitian and 
then multiply the solution by $D^{\dagger}$.
The simple conjugate gradient algorithm reads
\begin{algorithmic}
\STATE $\psi \gets \psi_0$
\STATE $r \gets \eta - D \psi$
\STATE $p \gets r$
\WHILE {$|r| \geq r_{min}$} 
	\STATE $r_{old} \gets |r|$
        \STATE $\alpha \gets \frac{r_{old}}{|D^{\dagger} p|}$
        \STATE $\psi \gets \psi + \alpha p$
	\STATE $r \gets r - \alpha D D^{\dagger} p$

	\STATE $\beta \gets \frac{|r|}{r_{old}}$
	\STATE $p \gets r + \beta p$
\ENDWHILE
\label{alg. cg}
\end{algorithmic}

The compute intensive part of the algorithm is the multiplication by $D^{\dagger}D$, out of which the most elementary computational block is the evaluation of the single stencil, i.e. the right hand side of Eq.\eqref{eq. dirac op} for a given value of the index $n$. This involves eight $U$ matrices and eight spinor fields from the neighboring lattice sites, which corresponds to 1536 input bytes. The $U \times \psi$ matrix-vector multiplications require 1152 floating point operations for complex additions and multiplications.

\begin{figure}
\centering
\includegraphics[width=180pt]{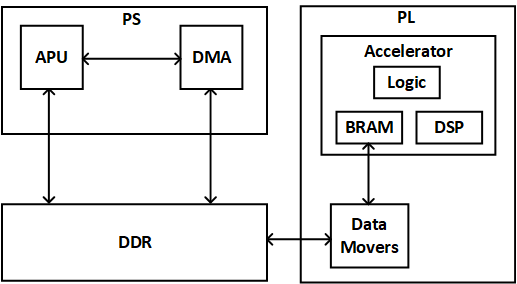}
\hspace{60pt}
\includegraphics[width=180pt]{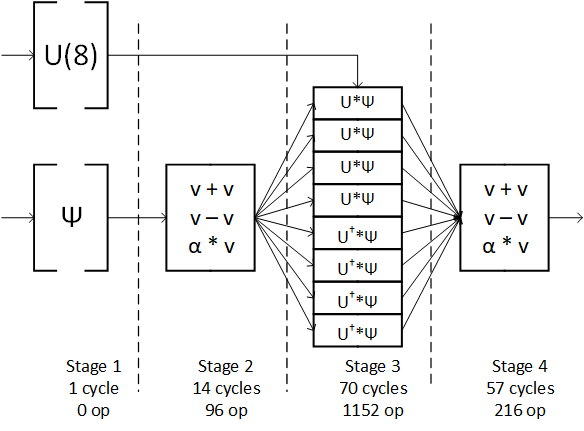}
\caption{Left panel: schematic view at the key components of Zynq MPSoC used to transfer data between PS and PL. Right panel: computation sequence of the stencil solver}
\label{soc_scm}
\end{figure}

A schematic view of the modern FPGA device is shown on the left panel of figure \ref{soc_scm}. As already mentioned it contains a processing part (PS) which in our case is equipped with an ARM processor, and the PL part with the programmable logic. The figure also shows the connections to the external DDR memory. One of the key decisions from the programming point of view is the software partitioning between the PS and PL parts. In the case of the CG algorithm there exists a natural candidate to be accelerated in hardware, which is the vector-matrix multiplication: the multiplication of the spinor field $\psi$ by the Dirac operator $D^{\dagger} D$ in equation \ref{eq. dirac eq}. In such scenario the main part of the CG algorithm is executed on the PS part. After each iteration the new residuum vector is returned to the DDR where the ARM processor evaluates the vector scalar products and the stopping criterion.

In order to easily compare the performance with the standard CPU architectures we use exclusively double precision numbers. This induces a considerable increase in the cost, because large logic resources are required to handle 64 bit arithmetics. A single addition or multiplication takes 14 clock cycles on the FPGA device used in this study. Modern GPU implementations of similar algorithms \cite{Clark:2009wm} use various combinations of low precision steps (single or half) in order to increase performance. FPGA devices offer a much broader spectrum of available fix point and floating point arithmetics which can be tuned to optimize the performance.

One of the great advantages of the programmable logic implementation is that one can use abstract types in the accelerated function without any loss of performance. Once the required data blocks are copied from the DDR memory to the memory blocks of the device, data can be accessed in parallel. In particular the ordering of the real and imaginary parts in memory does not longer play a role, as all the appropriate reshufflings will be executed in hardware data routing. In practice this allow us to work entirely with abstract types \verb|complex|, \verb|su3_vector|, \verb|su3_matrix| and \verb|su3_spinor|, the latter being simply
\begin{verbatim}
struct su3_spinor {
public:
        su3_vector s[4];
}
\end{verbatim}

However, in order to allow the compiler to take advantage of the natural parallelism of FPGA devices it is crucial to store data in PL in such a way that independent data is located in separate BRAM blocks. This limitation corresponds to the fact that in a single PL clock cycle only one memory element can be read from the BRAM block. In the computation of a single stencil one needs eight different $U$ matrices and we store them separately with the help of HLS pragmas such as \verb|HLS ARRAY_PARTITION|.
Although this requires duplicating the amount of stored data, the gain is considerable as thanks to that the stencil evaluation can be fully pipelined. Full pipelining means achieving initiation interval of one clock cycle, i.e. the hardware block can accept new input data at each clock cycle. The fixed structure of our computations, i.e. execution of exactly the same computations for each lattice site, allows for fully a pipelined Dirac matrix multiplication, if only enough logic resources are available on the device. Once
the data is stored appropriately the compiler parallelizes and pipelines the computations as shown on the right panel of figure \ref{soc_scm} by the schematic view of the logical data movement during a single stencil evaluation. The evaluation can be naturally divided into 5 stages.

In the first stage all the necessary data is copied from the BRAM memory blocks to local registers (declared with the \verb|ARRAY_PARTITION complete| attribute). The process requires only one clock cycle to collect all needed data. Later, in stage 2 linear combinations of input data, 8 additions and 8 subtractions of \verb|su3_vector| type, are evaluated. They are all performed in parallel, taking 14 clock cycles. The most compute intensive stage 3 involves \verb|su3_matrix| and \verb|su3_vector| multiplications. In total 1152 operations on \verb|double| are performed. Complete parallelization allows to execute them in a 5-layer operation cascade taking in total $5*14=70$ cycles. 
Finally, at stage 4 all contributions are added up to the final result. Because of the dependencies between consecutive partial results this creates a 4-layer operation cascade, which in total takes $57=(4*14)+1$ clock cycles, 4 additions plus one data copy.

Overall, the kernel requires 142 clock cycles and a total of 1464 basic operations to compute the final result since the reception of the input data. The kernel is fully pipelined: i.e. it can accept new input data at each clock cycle and produce the results with latency of 142 cycles. 

The implementation is fully scalable. If enough compute resources are available on the device, a second instance of the hardware accelerated kernel can be instantiated. Because the computations for each lattice site are completely independent, the set of lattice sites could thus be divided into two parts and each of them could be associated to one of the kernels, reducing the total computation time by a factor of two.

\section{Benchmarks results}
\label{sec. benchmarks}

We compiled our implementation for two systems: Xilinx ZU9EG. This is the device, which we had available, and Xilinx XCVU13P which we do not have physical access to. We used HLS version 2018.2. We report on the number of resources used for the compilation of the entire project in table \ref{tab. resources}. The first four entries are generated for the larger Xilinx XCVU13P device and differ by the initiation interval shown in column 2. The latter can be controlled by HLS pragmas limiting the number of instantiated blocks. In the last row of the table we separate resource usage gathered from the Xilinx ZU9EG.
\begin{table}
\begin{center}
\caption{ Resource consumption versus parallel execution. \label{tab. resources}}
\begin{tabular}{|ccccccc|}
\hline
latency&interval&bram&	dsp&	ff $[10^6]$ &	lut $[10^6]$&	uram\\
\hline
142&	1&	508&	6960&	1.58&   0.99&	696\\
151&	2&	448&	4320&	0.97&	0.64&	696\\
151&	2&	428&	3480&	0.83&	0.57&	696\\
162&	4&	412&	1740&	0.47&	0.35&	696\\
\hline
250&	120&	1388 &	546 &	0.13 &	0.09 & $-$ \\
&	& (76\%)& (21\%)&	(24\%)& (34\%)& acc.\\
&	& (95\%)& (21\%)&	(31\%)& (40\%)& full\\
\hline
\end{tabular}
\end{center}
\end{table}
The resource usage shows that it is possible to generate a fully pipelined kernel capable of computing single stencil within 142 cycles and maintaining initiation interval of 1 clock cycle. Smaller devices do not have enough internal memory resources to properly distribute array elements for the kernel to compute new results at each clock cycle. Therefore, the memory usage is crucial while achieving full pipelining. On ZU9EG 76\% of memory is used (the rest has to be reserved for Data Movers) but only 21\% of DSP blocks, which means that most of the computing resources are left unused due to lack of parallel memory interfaces and duplicated data units.

The maximal problem size which was run on the Xilinx ZU9EG device was $V=6^3 \times 8$ because of the limited memory resources available on that device. Problem sizes 
which were compiled for the Xilinx XCVU13P device and fit into that device memory range from $V=8^4$ to $V=8^3 \times 12$. The consumption of compute resources does not change as the problem size increases, only the amount of used memory blocks increases.

The timings are collected from execution and compilation logs by inspecting the trace data captured during algorithm execution in hardware. We count the number of cycles actually needed for the execution of particular steps of the calculation (reading data in from DDR, evaluating all stencils, writing data out to DDR). Assuming that the input data does not need to be loaded from the DDR memory for each consecutive call of the accelerated function, the duration of the computations can be calculated knowing the following parameters: the interval $\delta$ and latency $\tau$ of the compute kernel, the clock frequency $\nu$, the number of lattice sites $V$ to be processed and the number of FLOPs per lattice site $f$,
\begin{equation}
\textrm{performance} = V \times f \times \nu / ( V \times \delta + \tau ) 
\end{equation}
In table \ref{tab. flops} we report on the obtained performances in GFLOPs. 
\begin{table}
\begin{center}
\caption{Performance in GFLOPs. \label{tab. flops}}
\begin{tabular}{|c|c|c|}
\hline
Interval & Clock frequency & Performance \\
& [MHz] & [GFLOPs] \\
\hline					
1&500&676\\
2&500&351\\
4&500&179\\
\hline
120&150&1.82\\
\hline
\end{tabular}
\end{center}
\end{table}
Maximum performance can be achieved with the design that has the lowest initiation interval. That is possible on the XCVU13P device with has enough memory and DSP block available. Assuming no data transfer is required and the clock frequency is at the level of 500 MHz (kernel timing closure obtained) the generated computing logic reaches 676 GFLOPs. The performance drops linearly with the amount of fully utilized kernels. 

Due to limited memory resources we have achieved only 1.8 GFLOPs on the available ZU09EG. Not only high number of clock cycles for initiation interval but also lower clock frequency (Virtex Ultrascale+ is the high-end devices family and contains URAMs) are the reason for slower design.

When the kernels have to wait until all data is transported from the DDR to PL, we achieved 1.3 GFLOPs on the available platform. The ratio between with and without data transport is strongly dependent on the hardware infrastructure available on the FPGA chip itself and the hardware platform it is mounted on. Our ZCU09EG device has 4 hardware interfaces between DDR and PL. Using Virtex Ultrascale+ devices, one can profit from tens of embedded multigigabit transceivers to stream data into the device. Another solution is to use a direct connection between a DDR memory die and the logic as in Ref.\cite{Janson2017HighlyPL}, where the data is streamed into the programmable logic block.

\section{Conclusion}
\label{sec. conclusions}

In this contribution we have presented our implementation of the Conjugate Gradient algorithm as used Lattice QCD for FPGA devices. Our hardware accelerated kernel is fully parallelized and pipelined. We benchmarked our solution on a relatively small Xilinx evaluation board. Compilation details for larger devices suggest that the obtained performance can comparable with the one obtained on modern CPU units. We conclude therefore that FPGA devices can be considered as a viable solution for HPC systems.

\section*{Acknowledgment}

This work was in part supported by Deutsche Forschungsgemeinschaft under Grant No. SFB/TRR 55
and by the polish NCN grant No. UMO-2016/21/B/ ST2/01492, by the Foundation for Polish Science grant no. TEAM/2017-4/39 and by the Polish Ministry for Science and Higher Education grant no. 7150/E-338/M/2018.

The project could be realized thanks to the support from Xilinx University Program and their donations.

\bibliographystyle{JHEP}
\bibliography{references}

\end{document}